# Extending the Concept of Analog Butterworth Filter for Fractional Order Systems

Anish Acharya[a], Saptarshi Das[b], Indranil Pan[c], and Shantanu Das[d]

a)  *Department of Instrumentation and Electronics Engineering, Jadavpur University, Salt Lake Campus, LB-8, Sector 3, Kolkata-700098, India.*
b)  *Communications, Signal Processing and Control Group, School of Electronics and Computer Science, University of Southampton, Southampton SO17 1BJ, United Kingdom.*
c)  *MERG, Energy, Environment, Modelling and Minerals ($E^2M^2$) Research Section, Department of Earth Science and Engineering, Imperial College London, Exhibition Road, London SW7 2AZ, United Kingdom.*
d)  *Reactor Control Division, Bhabha Atomic Research Centre, Mumbai-400085, India.*

**Authors' Emails:**

anishacharya91@gmail.com (A. Acharya)

saptarshi@pe.jusl.ac.in , s.das@soton.ac.uk (S. Das*)

indranil.jj@student.iitd.ac.in , i.pan11@imperial.ac.uk (I. Pan)

shantanu@magnum.barc.gov.in (Sh. Das)

**Abstract**
This paper proposes the design of Fractional Order (FO) Butterworth filter in complex *w*-plane ($w=s^q$; *q* being any real number) considering the presence of under-damped, hyper-damped, ultra-damped poles. This is the first attempt to design such fractional Butterworth filters in complex *w*-plane instead of complex *s*-plane, as conventionally done for integer order filters. Firstly, the concept of fractional derivatives and *w*-plane stability of linear fractional order systems are discussed. Detailed mathematical formulation for the design of fractional Butterworth-like filter (FBWF) in *w*-plane is then presented. Simulation examples are given along with a practical example to design the FO Butterworth filter with given specifications in frequency domain to show the practicability of the proposed formulation.

**Keywords:** Butterworth filter; fractional order filter; fractional order linear system; *w*-plane stability; analog filter design

## 1. Introduction

In recent years, fractional calculus is being widely used in modeling the dynamics of many real life phenomena due to the fact that it has higher capability of providing accurate description than the integer dynamical systems. This added flexibility of FO systems is mainly due to the fact that poles in higher Riemann sheets which often contribute to the important dynamical features of physical systems can easily be analyzed and modeled using this technique e.g. the consideration of hyper-damped and ultra-damped poles etc. Thus due to greater flexibility in modeling, more accurate analysis and control design, fractional derivatives have been useful in many fields like visco-elasticity, acoustics, rheology, polymeric chemistry, biophysics etc. [1]. The application of fractional order systems as design elements have mostly been used in control and signal processing community as FO controllers and FO filters [2], [3]. Though different fractional order controllers like CRONE, $PI^\lambda D^\mu$, FO lead-lag compensators have already been famous, there are very few attempts on systematic design of FO filters. Previous attempts on FO analog filter design have mostly been restricted in the generalization of the first order and second order analog Infinite Impulse Response (IIR) transfer function templates as fractional order transfer functions by Radwan *et al.* [4], [5]. But the authors in [4], [5] did not consider the pole locations and the nature of the poles in higher Riemann sheets which is inevitable while dealing with fractional order linear dynamical systems. Other relevant approaches of designing fractional second order filters can be found in Li *et al.* [6]. Initial effort in this direction can be seen in Soltan *et al.* [7] with fractional capacitor and inductor based Butterworth filter design. Our present paper is the first attempt to extend the concept of assigning poles of a fractional order Butterworth filter lying on a circle in complex *w*-plane with consideration of its stability using well known Matignon's theorem [8]. A detailed analysis regarding the stability of different fractional order systems can be found in Radwan *et al.* [9]. The poles of such a filter will comprise of several cases like co-existence of under-damped, hyper-damped and ultra-damped poles which has not been investigated yet. Several other approaches in fractional filter design may include step filter [10], [11], different sinusoidal oscillators [12], [13], multi-vibrator circuits [14] etc.

The Butterworth filter is one of the most popular analog filter design paradigms, first described in 1930 by Stephen Butterworth [15]. The basic philosophy of the conventional or integer order analog Butterworth filter is well practiced in various applications. It is designed to have a frequency response as flat as it is possible. The frequency response of these filters is monotonic and the sharpness of the roll-off from pass band to stop band is determined by the order of the filter. For conventional Butterworth filters the poles associated with the magnitude squared function are equally distributed in angle on a circle in the complex *s*-plane around the origin and having radius equal to the cut-off frequency ($\Omega_c$). When the cut-off frequency and the filter order are specified, the poles can be obtained readily and from the pole position the transfer function of the filter can easily be obtained. Now, while designing a Butterworth filter we generally have four specifications, the pass band frequency ($\Omega_p$), stop band frequency ($\Omega_s$), maximum allowable pass band and stop band attenuation ($\alpha_p, \alpha_s$). The motivation of designing a fractional order Butterworth filter instead of the conventional integer order Butterworth filter is highlighted in this paper in the light of optimally finding the filter order to meet these design specifications rather than designing an over-specified filter.

As discussed earlier, given these specifications when a Butterworth filter is to be designed, the optimum order which is capable of meeting these specifications may be a fraction which is conventionally raised to the nearest integer value due to unavailability of filter order being a fraction. This is just similar to adding a bit more factor of safety in the design or over-satisfying the design specifications as the standard practice was to opt for a filter with next integer value. Hence it is obviously desirable neither to under satisfy nor to over satisfy the design specifications which can easily be met with the concept of fractional order filters with tunable parameters. With further enhancement of the realization techniques for fractional order filters, it will be possible to satisfy the exact requirements according to the given specifications. The other reason behind opting for fractional order Butterworth filter over the integer order one is the flexibility of design with the poles in *w*-plane, as it is well known that the *s*-plane design is a sub-set of the fractional order design considering the possibility of existing filter poles in different Riemann sheets.

Although there are several works on Butterworth filter design in past but its modification with the concept of fractional order derivatives is new. By doing so, the conventional *s*-plane stability needs to be modified first for fractional systems and we here systematically proposed the way of designing such filters by considering the *w*-plane stability given by Matignon's theorem [8] for fractional order systems. In a recent paper Soltan *et al.* [7], the design of the fractional order Butterworth filter has been reported. However the approach adopted in [7] has been quite different from the one proposed in our present paper. In [7] the fractional Butterworth filter design has been attempted from a circuit theory point of view, where two passive elements have been taken as $Z_1 = s^\beta L$ and $Z_2 = 1/s^\alpha C$ and thus designed the Butterworth filter following the conventional circuit design principles. However, the number of stable poles differs from the expected one. For example, in the pole location data provided in [7], a 0.7 order filter having two poles on the circle has been demonstrated. However, it is understandable from [16], [17] that the filter order being 0.7 we would expect 7 poles distributed over 10 Riemann sheets. Thus the formulation of FO Butterworth filter has been presented in a different way in order to generalize the design technique with state of the art analysis methodologies concerning the stability of fractional order linear systems. It is to be noted that in [7], the number of poles has been considered to be four which is a special case of fractional second order Butterworth filter whereas many other variants may also exist and has been generalized and illustrated with design examples in the present paper.

In this paper the fractional Butterworth filter has been formulated in the *w*-plane ($w = s^q, q \in \mathbb{R}_+$). Thus all possible types of poles (under-damped, hyper-damped, ultra-damped) have been taken into account. An analytical formulation of the filter is done in the *w*-plane and certain conditions for the location of the stable and the unstable poles are derived. The essence of classical Butterworth filter is kept unchanged and it is found that even in case of the fractional order formulation, the poles are placed along the circumference of a circle whose radius is equal to the cut-off frequency. Thus, by the proposed formulation all the stable poles have been taken into account. The corresponding transfer functions are obtained in *w*-plane and are then mapped back into the *s*-plane using the relation $s = w^{1/q}$. The corresponding response curves are obtained to confirm the maximally flat Butterworth nature.

Since this is one of the first approaches towards filter design in the *w*-plane, some generic issues related to fractional order filter design need to be discussed. It is known from the elementary knowledge of fractional order linear systems that a filter having an order $P/Q$ implies that $P$ number of poles are distributed over $Q$ Riemann sheets. However, this result is a serious issue related to the order of the filter. Say the order of 0.9 has been wrongly estimated as 0.99 or 0.999 or that a first order system has been wrongly estimated as 1.001. So, it is worth considering whether increase in the number of poles and number of Riemann sheets for capturing delicate frequency domain information is actually required, since this would lead to higher design complexity [16], [17]. For a 0.9 order filter, the proposition will require 9 poles distributed over 10 Riemann sheets, however if the order is estimated to be 0.99, 99 poles distributed over 100 Riemann sheets are to be handled ,which will be cumbersome to realize in hardware. This resulting complexity would make it difficult to generalize filters in fractional domain and the advantage of greater flexibility will be outweighed by fabrication costs. So, a design scheme which considered the relative trade-off between complexity and accuracy has been adopted. Thus for a filter whose overall order is a fraction, it is better to realize it only up to the first decimal place (by truncating the trailing decimals) and then increasing the accuracy by intuitive judgment for improving the frequency domain characteristics.

There are many different connotations related to the term "filtering". In the classical sense, filter design refers to FIR or IIR structures which are designed to satisfy some frequency domain specifications like cut-off frequencies for the pass-band or stop band, maximum allowable ripples in magnitude response etc. This notion of "filtering" is considered in the present paper. The other notion of filtering may be from the point of view of the theory of linear estimation [18]. In such cases the filter design results in some FIR/IIR structure which minimizes some performance objective in time domain, e.g. Wiener filter. The third paradigm of filtering is a model based state estimation or parameter estimation technique like the Kalman filter, extended and unscented Kalman filter etc. This concept has been augmented to obtain different enhanced performance objectives and for varied complex systems with $H_\infty$ filtering and particle filtering [19]. In [20], discusses the latest development in this category using fuzzy logic. This has been an active research topic for the last few decades. But in recent years, with the advent of fractional calculus in system design, our notions of filtering for all the three above mentioned paradigms need to be modified in this light. Our present work can be seen as a first step for systematic fractional filter design with the consideration of stability for fractional order systems. Time domain filters like fuzzy [20] or $H_\infty$ filtering [21] require some sort of a system model. However, in case of conventional IIR/FIR filtering, the technique is applied on the frequency spectrum of the signal directly. Hence the need of a system model is eliminated.

In cases where the signal is corrupted by random disturbances or noise, linear estimation based filtering technique works well [18]. But in many industrial processes there is a clearly identified frequency band of interest (or pass band), and rest of the frequency spectrum is not important. In such cases, simple FIR/IIR filtering techniques are preferred over model based filters to reduce design complexity. All these stochastic filtering techniques require additional parameters for their design unlike classical FIR/IIR filter design. For example, mixing properties (additive/multiplicative nature), statistical properties of the noise before mixing, parameterized linear/nonlinear model of the system generating the signal, upper bound of the disturbance spectra etc. are required depending on the specific design techniques.

We propose here a new paradigm of designing fractional Butterworth filter to meet exact design requirements which is not possible in conventional integer order filter design. We have also shown that integer order Butterworth filter will always under-satisfy or over-satisfy the design specifications. The proposed technique is a new way of looking at analog filter design with fractional calculus based techniques. The magnitude roll-off can be adjusted to any desired slope, which can only be achieved using fractional order systems. While dealing with fractional order transfer functions, the *w*-plane stability concepts or Matignon's theorem should be used instead of the conventional *s*-plane stability. It is to be noted that in recent years several efforts have been there to propose new automatic controller design using the concept of *w*-plane stability for example [22]-[24]. This is the first attempt to extend the concept of stability in automatic control design to analog Butterworth filter design using *w*-plane stability notion for FO systems. Existing state of the art design methods would increase the order of the IIR filter to meet design specifications or to get a smaller transition band. Since fractional order systems are inherently infinite dimensional, they naturally model the exact magnitude roll-off. The order of realization can be chosen by the user after the exact order of the filter is calculated from the frequency domain specifications.

The rest of the paper is organized as follows. Section 2 reviews the fundamentals of fractional calculus required to understand the mathematical formulation of the fractional order Butterworth filter. In Section 3 the mathematical formulation for the realization of fractional Butterworth filter is introduced. In Section 4 simulation results are presented. Section 5 illustrates a realistic example of designing a Fractional order Butterworth Filter from given specifications. The paper ends in Section 6 with the conclusions and some comments about the future scope in this field.

## 2. Basics of fractional calculus based linear dynamical systems

The basics of fractional calculus that has been used in this paper, are discussed here as the background of the filter design technique.

### 2.1 Functions used in fractional calculus

Some important functions in relation to fractional calculus are as follows:

i. *Gamma function*

$$\Gamma(z) = \int_0^\infty e^{-u} u^{z-1} du, \forall z \in \mathbb{R} \tag{1}$$

For complex $z$ the real part has to be finite to get a finite value of the gamma function.

ii. *Beta function*

$$B(p,q) = \int_0^1 (1-u)^{p-1} u^{q-1} du, \, p,q \in \mathbb{R}_+ \tag{2}$$

The relation between the gamma function and the beta function is $B(p,q) = \dfrac{\Gamma(p)\Gamma(q)}{\Gamma(p+q)}$.

### 2.2 Definitions of fractional differ-integrals

The two popular definitions of fractional differ-integration are as follows [1]

*i. Grunwald-Letnikov (G-L) Definition:*

This is basically an extension of the backward finite difference formula for successive differentiation. This formula is used widely for the numerical solution of fractional differentiation or integration of a function. According to Grunwald-Letnikov definition the $\alpha^{th}$ order differ-integration of a function $f(t)$ is defined as

$$D_t^\alpha f(t) = \lim_{h \to 0} \frac{1}{h^\alpha} \sum_{j=0}^{\infty} (-1)^j \frac{\Gamma(\alpha+1)}{\Gamma(j+1)\Gamma(\alpha-j+1)} f(t-jh) \qquad (3)$$

*ii. Riemann-Liouville (R-L) Definition:*

This is an extension of *n*-fold successive integration. The $\alpha^{th}$ order integration of a function $f(t)$ is defined as

$$_aI_t^\alpha f(t) = \frac{1}{\Gamma(-\alpha)} \int_a^t \frac{f(\tau)}{(t-\tau)^{\alpha+1}} d\tau \text{ for } a, \alpha \in \mathbb{R}, \alpha < 0 \qquad (4)$$

and similarly the $\alpha^{th}$ order differentiation of a function $f(t)$ is defined as

$$_aD_t^\alpha f(t) = \frac{1}{\Gamma(n-\alpha)} \frac{d^n}{dt^n} \int_a^t \frac{f(\tau)}{(t-\tau)^{\alpha-n+1}} d\tau \text{ for } n-1 < \alpha < n \qquad (5)$$

## 2.3 Stability regions in *w*-plane

Generally, for the multi-valued function defined as $w = s^{\frac{1}{Q}} = s^q$ where $Q \in N$ there are $Q$ sheets in the Riemann surface. The sector $-\pi/Q < \arg(w_k) < \pi/Q$ corresponds to the first Riemann sheet and $k$ is the number of poles. Thus when poles of any fractional order system in *s*-plane are mapped into *w*-plane, all the Riemann sheets are taken into account. Hence all the different kinds of poles showing different dynamical behaviors are considered, especially the ultra-damped and hyper-damped poles [25]. The stable region is now constrained within the region [8]

$$\arg|(w_k)| > q\frac{\pi}{2} \qquad (6)$$

Table 1 shows the different pole types and their corresponding regions in the *w*-plane.

**Table 1: Type of poles depending on their position in the *w*-plane**

| Pole position | Comment on pole nature |
|---|---|
| $|\arg(w)| < q\pi/2$ | Unstable |
| $q\pi/2 < |\arg(w)| < q\pi$ | Stable under-damped |
| $q\pi < |\arg(w)| < \pi$ | Stable hyper-damped |
| $|\arg(w)| = \pi$ | Stable ultra-damped |

## 3. Proposal for a new fractional order Butterworth (FBW)-like filter

In this section, a fractional Butterworth like filter has been proposed using the concept of $w$-plane stability. The poles are located on the $w$-plane which consists of all the poles lying on different higher Riemann sheets. Then the stability criterion is taken under consideration and all the unstable poles are rejected and a fractional order BW like filter can be obtained. The concept is somewhat similar to the conventional notion of finding integer BW filter poles lying on a circle and then discarding the right-half plane poles in order to avoid instability. The FBW filter is developed on the lines of the integer order Butterworth filter with the considerations of the stability of fractional linear systems [26].

*Theorem 3.1:*

The pole locations of a FO Butterworth low-pass filter of any real order $P/Q, \{P,Q\} \in \mathbb{Z}_+$ are given by $w_k = \pm j\Omega_C e^{j(2k-1)\pi/2P}$, where $k = 1, 2, \cdots, P$ and the condition for stability of the roots is given by $\arg|(w_k)| > \frac{q\pi}{2}$, where $q = 1/Q$ is the commensurate order and $P$ is the number of poles distributed over $Q$ Riemann sheets.

*Proof:*

The squared magnitude function of an analog Butterworth filter is given by the form

$$|H(j\Omega)|^2 = \frac{1}{1+\left(\frac{\Omega}{\Omega_c}\right)^{2N}} \qquad (7)$$

where, $N$ is the order of the filter.

Now considering $s = j\Omega \Rightarrow \Omega = s/j$, the squared magnitude system function is of the form

$$H(s)H(-s) = \frac{1}{1+\left(\frac{s}{j\Omega_c}\right)^{2N}} \qquad (8)$$

For fractional order filters the magnitude squared function would be written in terms of fractional powers of Laplace variable $s^q = w$. Then we have the magnitude squared function in $w$-plane as

$$H(w)H(-w) = \frac{1}{1+\left(\frac{-w^2}{\Omega_C^2}\right)^P} \qquad (9)$$

So if $P/Q, \{P,Q\} \in \mathbb{Z}_+$ be the order of the filter where $q = 1/Q, q \in \mathbb{R}_+$, where $q$ is the commensurate order hence we now have $P$ number of poles distributed over $Q$ Riemann sheets. Therefore the design of $P/Q$ order FBW like filter is similar to design a $P^{th}$ order integer

Butterworth filter with the poles equally distributed on the circumference of a circle in the complex *w*-plane while only discarding the unstable poles using the relation (6). So in $w$-plane we find the pole locations from the characteristic polynomial

$$1 + \left(\frac{-w^2}{\Omega_C^2}\right)^P = 0 \tag{10}$$

This is being proposed as the Butterworth like polynomial which has been expressed as a function of *w* instead of *s*. Hence the pole locations are found to be $w_k = \pm j\Omega_C e^{j(2k-1)\frac{\pi}{2P}}$, where $k = 1, 2, \cdots, P$. and the criterion for stable poles is the same as (6). □

The results are illustrated and the natures of the poles are investigated for various cases. Let $P = 1$ then the poles are located at $w_k = \pm j\Omega_C e^{j\frac{\pi}{2}} = \pm \Omega_C$. The poles located within the region $\pm q\frac{\pi}{2}$ are unstable, hence it is obvious that $w = +\Omega_C$ corresponds to unstable system. Therefore, we consider only $w = -\Omega_C$ and thus the transfer function of $(1/Q)^{th}$ order filter and its mapping back to the *s*-plane is found to be

$$H(w) = \frac{\Omega_C}{w + \Omega_C} \Rightarrow H(s) = \frac{\Omega_C}{s^q + \Omega_C} \tag{11}$$

Therefore, here we got a single ultra-damped pole for the FBW filter.

Now when $P = 2$ is considered, the pole locations are found to be $w = \pm j\Omega_C e^{j\frac{\pi}{4}}$, $\pm j\Omega_C e^{j\frac{3\pi}{4}}$ which may be stable depending on the commensurate order $q$. Therefore, the transfer function becomes

$$H(s) = \frac{\Omega_C^4}{(s^{2q} - \sqrt{2}\Omega_C s^q + \Omega_C^2)(s^{2q} + \sqrt{2}\Omega_C s^q + \Omega_C^2)} \tag{12}$$

It is interesting to note that for classical second order Butterworth filter, seems to have the same structure as in (12), except that the unstable poles are in the right half *w*-plane (yielding negative coefficient of transfer function). For similar classical design i.e., $q = 1$, the term $(s^{2q} - \sqrt{2}\Omega_C s^q + \Omega_C^2)$ used to be discarded as they represent unstable complex conjugates in *s*-plane. But for fractional order systems, these poles may be stable and thus must be included in the design depending on the respective stability criteria. Therefore fractional order transfer functions with negative coefficients with poles lying in the right half part of *w*-plane, but outside the instability region, is an effective way for the development of filters having any arbitrary real order.

Now the nature of the poles is dependent on the commensurate order as discussed in section 2.3. On analyzing the nature of the roots for different combinations of $P$ and $q$, i.e. for different orders of the filters, the nature of the poles need to be judged for possible effects in the filtering action using the notions given in Table 1. For simplicity, the pole locations of the fractional filter are designed for overall order less than unity. In other words, for a fractional Butterworth like filter with order greater than unity, the integer and the fractional parts are to be designed separately and then needs to be cascaded together. This is important due to the fact that for order greater than unity (e.g. order $N + \frac{P}{Q}, N \in \mathbb{Z}_+, Q > P, \{P,Q\} \in \mathbb{R}_+$), the number of poles in w-plane increases (i.e. $(NQ+P)$ number of poles distributed over $Q$ Riemann sheets). Therefore, it is expedient to design the $N^{th}$ order BW filter using conventional notion of integer order BW filter and then designing the $(P/Q)^{th}$ order filter using the proposed technique. It is important to mention that there should not be any common factor in $\{P,Q\}$, in order to avoid unnecessary increase in the number of poles and number of Riemann sheets. The design approach presented here is valid for fractional filters having any arbitrary real order, represented by a rational fraction less than unity with no common factor among the numerator and denominator. This is because the generalization of finding unstable poles among the $(NQ+P)$ number of poles is cumbersome.

In Table 2 the pole locations considering $\Omega_c = 1$ rad/sec and $Q > P, \forall Q = [1, 2, \cdots, P-1]$ has been reported. As argued above, the real parts of the poles are positive in few cases, though they may represent stable systems depending on the value of $Q$ which determines the angle of the unstable region. Therefore it can be seen that for a fractional filter of order $(P/Q) < 1$, $P$ represents the number of poles in the complex w-plane irrespective of the value of $Q$ and the $Q$ plays the role of choosing only the stable poles by defining the instability region using relation (6). From Table 2 it can be seen that the filter will have combined dynamics of under-damped, hyper-damped poles and ultra-damped poles.

Table 2: Pole locations for different fractional order Butterworth polynomials in w-plane

| Q | Unstable region | P | Stable pole locations |
|---|---|---|---|
| 2 | $\|\arg(w_k)\| > \frac{\pi}{4}$ | 1 | -1 |
| 3 | $\|\arg(w_k)\| > \frac{\pi}{6}$ | 1 | -1 |
| | | 2 | -0.7071+0.7071i, -0.7071-0.7071i, 0.7071-0.7071i, 0.7071+0.7071i |
| 4 | $\|\arg(w_k)\| > \frac{\pi}{8}$ | 1 | -1 |
| | | 2 | -0.7071+0.7071i, -0.7071-0.7071i, 0.7071-0.7071i, 0.7071+0.7071i |

| | | 3 | -0.5000+0.8660i, -1, -0.5000-0.8660i, 0.5000-0.8660i, 0.5000+0.8660i |
| --- | --- | --- | --- |
| | | 1 | -1 |
| | | 2 | -0.7071+0.7071i, -0.7071-0.7071i, 0.7071-0.7071i, 0.7071+0.7071i |
| 5 | $\left|\arg(w_k)\right| > \dfrac{\pi}{10}$ | 3 | -0.5000+0.8660i, -1, -0.5000-0.8660i, 0.5000-0.8660i, 0.5000+0.8660i |
| | | 4 | -0.3827+0.9239i, -0.9239+0.3827i, -0.9239-0.3827i, -0.3827-0.9239i, 0.3827-0.9239i, 0.9239-0.3827i, 0.9239+0.3827i, 0.3827+0.9239i |

In Table 3 the stable fractional order transfer functions of the Butterworth like filter in $w$-plane is shown in terms of the cut-off frequency $\Omega_c$. For fractional order filter design, the specifications on $\Omega_c$ should be suitably transformed so as to match the $s \leftrightarrow w$ transformation. If it be considered that the circle having a radius as the cut off frequency $\Omega_c$ in $s$-plane, then using the relation $w = s^q$ the transformed radius ($\overline{\Omega_c} = \Omega_c^q$) in $w$-plane becomes

$$\left.\begin{array}{l}\overline{\Omega_c} < \Omega_c \text{ for } \Omega_c > 1 \\ \overline{\Omega_c} > \Omega_c \text{ for } \Omega_c < 1\end{array}\right\} \forall q \in \mathbb{R}_+, q \in (0,1) \qquad (13)$$

**Table 3: Stable fractional order transfer functions using the concept of Butterworth filter in w-plane**

| $P$ | $H(s)$ |
| --- | --- |
| $P = 1$ | $\dfrac{\overline{\Omega_c}}{s^q + \overline{\Omega_c}}$ |
| $P = 2$ | $\dfrac{\overline{\Omega_c}^4}{(s^{2q} - \sqrt{2}\overline{\Omega_c}s^q + \overline{\Omega_c}^2)(s^{2q} + \sqrt{2}\overline{\Omega_c}s^q + \overline{\Omega_c}^2)}$ |
| $P = 3$ | $\dfrac{\overline{\Omega_c}^5}{(s^q + \overline{\Omega_c})(s^{2q} - \overline{\Omega_c}s^q + \overline{\Omega_c}^2)(s^{2q} + \overline{\Omega_c}s^q + \overline{\Omega_c}^2)}$ |
| P=4 | $$\frac{\overline{\Omega_c}^8}{(s^{2q}-0.7654\overline{\Omega_c}s^q+\overline{\Omega_c}^2)(s^{2q}+0.7654\overline{\Omega_c}s^q+\overline{\Omega_c}^2)(s^{2q}+1.848\overline{\Omega_c}s^q+\overline{\Omega_c}^2)(s^{2q}-1.848\overline{\Omega_c}s^q+\overline{\Omega_c}^2)}$$ |
| P=5 | $$\frac{\overline{\Omega_c}^9}{(s^{2q}+1.618\overline{\Omega_c}s^q+\overline{\Omega_c}^2)(s^{2q}+0.618\overline{\Omega_c}s^q+\overline{\Omega_c}^2)(s^{2q}-1.618\overline{\Omega_c}s^q+\overline{\Omega_c}^2)(s^{2q}-0.618\overline{\Omega_c}s^q+\overline{\Omega_c}^2)(s^q+\overline{\Omega_c})}$$ |
| P=6 | $$\frac{\overline{\Omega_c}^{12}}{\begin{array}{c}(s^{2q}+1.932\overline{\Omega_c}s^q+\overline{\Omega_c}^2)(s^{2q}+1.414\overline{\Omega_c}s^q+\overline{\Omega_c}^2)(s^{2q}+0.5176\overline{\Omega_c}s^q+\overline{\Omega_c}^2)\\(s^{2q}-1.414\overline{\Omega_c}s^q+\overline{\Omega_c}^2)(s^{2q}-1.932\overline{\Omega_c}s^q+\overline{\Omega_c}^2)(s^{2q}-0.5176\overline{\Omega_c}s^q+\overline{\Omega_c}^2)\end{array}}$$ |
| P=7 | $$\frac{\overline{\Omega_c}^{13}}{\begin{array}{c}(s^{2q}+1.802\overline{\Omega_c}s^q+\overline{\Omega_c}^2)(s^{2q}+1.247\overline{\Omega_c}s^q+\overline{\Omega_c}^2)(s^{2q}+0.445\overline{\Omega_c}s^q+\overline{\Omega_c}^2)(s^{2q}-1.802\overline{\Omega_c}s^q+\overline{\Omega_c}^2)\\(s^{2q}-1.247\overline{\Omega_c}s^q+\overline{\Omega_c}^2)(s^{2q}-0.445\overline{\Omega_c}s^q+\overline{\Omega_c}^2)(s^q+\overline{\Omega_c})\end{array}}$$ |
| P=8 | $$\frac{\overline{\Omega_c}^{16}}{\begin{array}{c}(s^{2q}+1.962\overline{\Omega_c}s^q+\overline{\Omega_c}^2)(s^{2q}-1.962\overline{\Omega_c}s^q+\overline{\Omega_c}^2)(s^{2q}+1.663\overline{\Omega_c}s^q+\overline{\Omega_c}^2)(s^{2q}-1.663\overline{\Omega_c}s^q+\overline{\Omega_c}^2)(s^{2q}+1.111\overline{\Omega_c}s^q+\overline{\Omega_c}^2)\\(s^{2q}-1.111\overline{\Omega_c}s^q+\overline{\Omega_c}^2)(s^{2q}+0.3902\overline{\Omega_c}s^q+\overline{\Omega_c}^2)(s^{2q}-0.3902\overline{\Omega_c}s^q+\overline{\Omega_c}^2)\end{array}}$$ |
| P=9 | $$\frac{\overline{\Omega_c}^{17}}{\begin{array}{c}(s^{2q}+1.879\overline{\Omega_c}s^q+\overline{\Omega_c}^2)(s^{2q}-1.879\overline{\Omega_c}s^q+\overline{\Omega_c}^2)(s^{2q}+1.532\overline{\Omega_c}s^q+\overline{\Omega_c}^2)(s^{2q}-1.532\overline{\Omega_c}s^q+\overline{\Omega_c}^2)\\(s^{2q}+\overline{\Omega_c}s^q+\overline{\Omega_c}^2)(s^{2q}-\overline{\Omega_c}s^q+\overline{\Omega_c}^2)(s^{2q}+0.3472\overline{\Omega_c}s^q+\overline{\Omega_c}^2)(s^{2q}-0.3472\overline{\Omega_c}s^q+\overline{\Omega_c}^2)(s^q+\overline{\Omega_c})\end{array}}$$ |

In other words, the radius of the circle on which the poles of the Butterworth filter lies, gets reduced or amplified depending on the cutoff frequency being higher or lower than unity respectively. In Table 3 the transfer functions have been expressed in terms of the transformed radius $\overline{\Omega_c}$ as the poles are those belonging to the *w*-plane, including the higher Riemann sheets. Similar to the conventional integer $P^{th}$ order Butterworth filter transfer function, *P* is varied from 1 to 9. These transfer functions have a similar template to the original Butterworth filter, but have the following differences:

   i. *s* is replaced by $w = s^q$
   ii. $\Omega_c$ is replaced by $\overline{\Omega_c} = \Omega_c^q$
   iii. Integer BW poles are located only in the left half *s*-plane. For fractional BW like filter, same set of poles come in the right half *w*-plane as a mirror image with respect to the imaginary *w*-axis.

### 3.1 Remarks

Though the above mentioned formula is expected to be quite generalized however there is a small issue to be taken into account. Firstly, this is a trade-off design, i.e. truncation of the higher decimal places is needed to reduce complexity of fractional filter realization because otherwise placement of a large number of poles and deriving their locations in different Riemann sheets cannot be handled with ease. Therefore, we truncated the filter order to make a design with only up to the first decimal place and then improving the accuracy by intuitive judgment for improving frequency domain characteristics.

Secondly, another thing that is to be noted is the fact that the above mentioned theorem will not hold true in case of filter order being an irrational fraction. For example an overall order of $\sqrt{3}$ cannot be realized using the above mentioned theorem. In such cases a truncated version of the recurring decimal needs to be considered.

## 4. Simulations and results

Here different commensurate orders have been chosen and correspondingly the number of poles distributed over different number of Riemann sheets has also been considered to be different. The nature of the poles for all those cases is investigated. Initially the commensurate order is chosen to be $Q=10$ and the order of the respective filters has been obtained by varying $P$ and thus considering all possible filter orders from 0.1 to 0.9 while the location and the nature of the poles are also studied accordingly. By nature of the poles, it is implied how many poles are stable, among the stable poles how many lie in the under-damped, hyper-damped or ultra-damped regions. Then simulations are added in order to verify the behavior of the FBWF designed using the proposed mathematical formulation.

Similarly all possible cases for different commensurate orders $Q=2$ to 9 have also been taken into account and correspondingly all possible different fractional orders of the Butterworth filter are studied and then simulated to get an overall idea about their response to unit step excitation. Using theorem 3.1 the simulations for time and frequency responses of different filters have been done and reported here, after discarding the unstable poles and therefore considering only the stable ones in the filter transfer function.

### 4.1 Comparison of frequency response of fractional Butterworth filters

Figure 1 shows the frequency response of the FBWF for various combinations of $P, Q \; \forall P = \{1, 2, \cdots, Q-1\}$. From these four set of frequency responses it can be observed that the magnitude and the phase roll-offs decrease with the increase in the value of $Q$. Similarly as $P$ increases (and hence the order of the FBWF), the magnitude and phase roll-off increases, for same value of $Q$. However the magnitude and phase roll offs are not as fast as the integer order filters. This typical phenomenon helps to meet design specifications precisely and obtain an accurate filtering action as required. In the simulation examples, the term 'order' of the FBWF represents the value $P$, determining the number of poles in the $w$-plane, which is independent of the commensurate order $Q$.

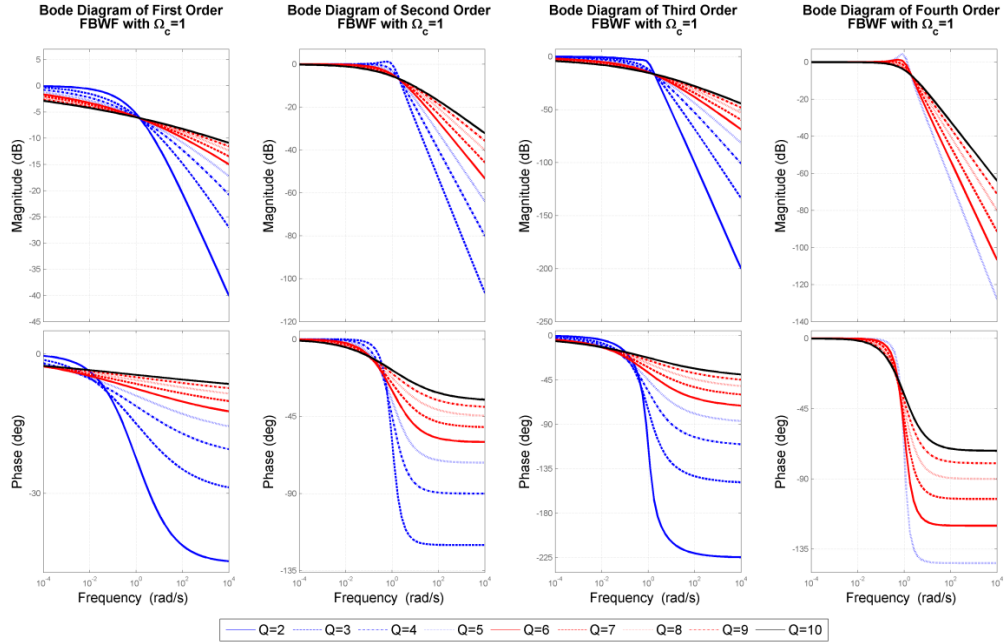

**Figure 1: Bode diagram of various orders of Fractional Butterworth like Filters with unit cut-off frequency for different values of Q.**

### 4.2 Step Response comparison

Figure 2 shows the time response of the fractional order BW filters whose Bode diagrams are shown in Figure 1. In second order FO BW filter due to the co-existence of under-damped and hyper-damped poles made the curve to have oscillatory behavior for $Q = 3$. Whereas, in third order FBWF the co-existence of 2 under-damped, 2 hyper-damped and 1 critically damped pole made the step response to have local oscillations (due to the under-damped pole) but not let the step response magnitude exceed unity at all times. As we know that such ultra-damped poles cause extreme sluggishness in the step response [25] but exhibits local oscillations instead of a monotonic step response due to having two under-damped poles for $Q = 3$. Similarly for the fourth order FBWF, co-existence of two under-damped poles and many hyper-damped poles makes the system oscillatory for $Q = 3$ to $7$. For further higher values of $Q$, since the instability region and hence the under-damped region of the primary Riemann sheet becomes more narrow, as a result all the poles lie in the hyper-damped region indicating a sluggish step response.

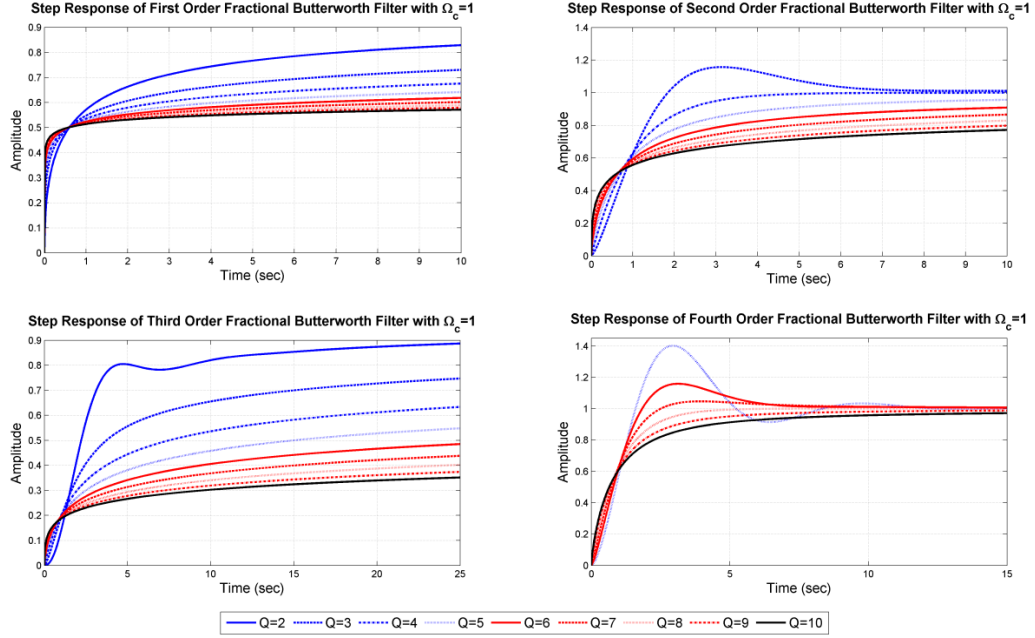

**Figure 2: Step response of various orders of fractional Butterworth like filter with unit cut-off frequency for different values of Q.**

### 4.3 Pole locations of the FBW filter in complex *w*-plane

Figure 3 shows the pole locations of the FBW filter in the *w*-plane for different commensurate orders. The first order ($P=1$) FBWF has one ultra-damped pole and is always stable irrespective of the value of $Q$. The presence of the ultra-damped pole implies a very sluggish time response when the filter is subjected to a unit step input. A second order ($P=2$) FBWF has two under-damped and two hyper-damped poles which are always stable for all integer values of $Q$. It can be seen that two of the poles lie on the right half *w*-plane, but they are still outside the angular sector of instability $\pm q\pi/2$. The third order FBWF filter has two under-damped, two hyper-damped and one ultra-damped poles. Similar to the previous case, it is always stable for all integer values of $Q$. The fourth order FBWF has four under-damped and four hyper-damped poles. However unlike the previous cases, the transfer function given in Table 3 is only stable for $Q>3$. For $Q=2$ and $Q=3$, one unstable pair of poles should be excluded to obtain the filter transfer function. The fifth order case has four under-damped, four hyper-damped poles and one ultra-damped pole. It is stable for $Q>2$. For $Q=2$, the one unstable pair of poles must be excluded to obtain a stable filter transfer function. For the sixth order case six under-damped and six hyper-damped poles are there. The filter is stable for $Q>5$. For Q=2 to 5, the one unstable pair of poles must be excluded to obtain the filter transfer function. The seventh order case has six under-damped, six hyper-damped poles and one ultra-damped pole. The filter is stable for $Q>3$. For $Q=2$ and $Q=3$, the one unstable pair of poles must be excluded to obtain the filter transfer function. In the eighth order case, there are eight under-damped and eight hyper-damped poles. The filter is stable for $Q>7$. For Q=2 to 5, the one unstable pair of poles must be excluded to obtain the filter transfer function. It is also observed that another pair of poles falls within the instability region for $Q = 2$. Thus for the eighth order FBWF with $Q = 2$, two pairs of unstable poles have to be excluded and the resulting transfer function has to be obtained. For ninth order BW filter a much simpler case can be observed where only one set of unstable poles have to be excluded for $Q=2$ to $4$. This filter has the same number of under-damped and hyper-damped poles for $Q = 2$ as in the eighth order case, in addition with an ultra-damped pole. It is always stable for $Q > 4$. It is to be noted that odd order of FBWF yields an ultra-damped pole which produces high sluggishness in the step response, whereas, with the even orders such oscillations are avoided. As mentioned above, it can be seen that for higher values of $P$ (higher than 3), some of the conjugate pairs are unstable for low $Q$ but stable for other higher values of $Q$. In such cases the stability check using (6) must be done and the unstable pair of poles needs to be discarded.

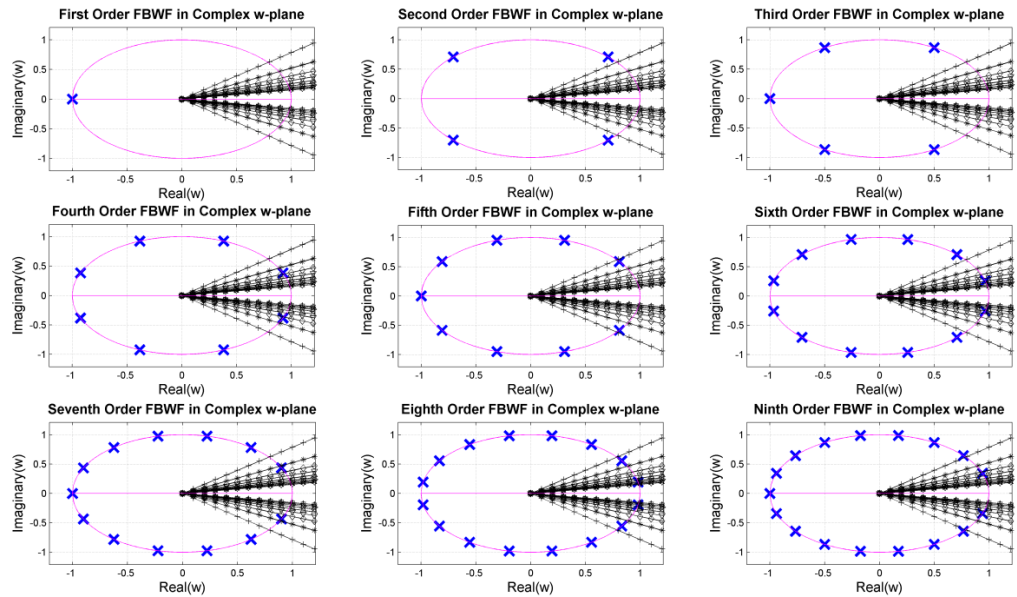

**Figure 3: Pole zero map of various orders of Butterworth filter in complex w-plane for different Q values.**

It is also interesting to note that the poles of the fractional order filter are equi-spaced in complex *w*-plane and also stable which has not been done by contemporary researchers. As conventionally followed in the fractional order control community, we must consider the *w*-plane stability notion or Matignon's theorem to verify that the poles of the filter fall in the stable region and also are equally distributed in *w*-plane.

## 5. Design advantages of fractional Butterworth filter

Generally, when a filter is to be designed it is expected to meet certain frequency domain design specifications. There are four major specifications to be dealt with, while designing an analog Butterworth filter. The frequency for the end of the pass band ($\Omega_p$) and the start of the stop band ($\Omega_s$) are specified and hence the starting and ending frequencies of the transition band are also specified beforehand. Then the maximum pass band attenuation ($\alpha_p$) and minimum stop band attenuation ($\alpha_s$) are also specified. In order to design a BW filter which meets all these specifications, the required order of the filter needs to be chosen first [26]. It can easily be obtained that given these specifications the exact order of the filter in terms of these specifications is given as equation (14)

$$N = \frac{\log\sqrt{\frac{10^{0.1\alpha_s}-1}{10^{0.1\alpha_p}-1}}}{\log\frac{\Omega_s}{\Omega_p}} \qquad (14)$$

Also, the cut-off frequency ($\Omega_c$) can be calculated from the specification of the stop-band ($\Omega_s$) using the following relation.

$$\Omega_c = \frac{\Omega_s}{\left(10^{0.1\alpha_s}-1\right)^{1/2N}} \qquad (15)$$

Obviously it can be deduced that the value of $N$ cannot always take integer values. In most cases it is more likely to take fractional values. So in the traditional integer order case, we had to over satisfy the design needs. For example, if a filter of order 1.5 was required in order to exactly meet the given specifications, a second order filter had to be designed. This can be overcome by fractional filter of an order exactly equal to 1.5. As mentioned before in order to design a filter of order 1.5, a 0.5 order filter can be cascaded with a first order filter. One realistic design example is illustrated here and supported with MATLAB based simulation studies.

### 5.1 Realistic Design example

Suppose given the specifications that the maximum pass band attenuation be 0.5dB and minimum stop band attenuation be 20 dB. The pass band and stop band edge frequencies are given as 2 rad/sec and 3 rad/sec. It is desired to realize a fractional order Butterworth like filter using the proposed technique.

Hence as from the given specifications $\alpha_p = 6\,dB$, $\alpha_s = 20\,dB$, $\Omega_p = 2\,rad/\sec$, $\Omega_s = 3\,rad/\sec$ [26]. Thus the desired order of the filter required is $N = 4.3195$. Truncating up to one decimal place the order of the filter can be approximated as $N = 4.3$. But from the convention of integer order BW filters, we had to over satisfy the requirements and design a filter of order 5. But using the proposed concept of FBW filter, we can now design a BW filter having an integer and a fractional part, having an exact order of 4.3. In the design process, the nominal fractional part has been represented as a ratio of two integers with no common factors i.e. $P/Q = 3/10$ for our case. Here it is worth noticing that a filter of order 4.3 could have been designed which has 43 poles distributed over 10 Riemann sheets. But in such cases the number of unstable poles would have increased and as such no generalization can be possible like that reported in Table 2 and Table 3. Therefore in order to keep simplicity, it is preferable that the nominal integer and fractional parts are designed separately using the respective concepts of designing BW polynomial in corresponding *s*-plane or *w*-plane and construct filters using stable poles only and then cascade these two parts. Here, we have taken a similar approach and put a FBWF of order 0.3 in cascaded with a 4$^{th}$ order conventional BW filter.

For the integer order part ($N = 4$) it is found using (15) that $\Omega_c^{N=4} = 1.6891$ rad/sec. For conventional integer order design the next integer value is used i.e. $N = 5$ and in that case the cut-off frequency $\Omega_c^{N=5} = 1.8948$ rad/sec. For integer order design the conventional 4$^{th}$ and 5$^{th}$ order Butterworth filter transfer functions are

$$H_{N=4}(s) = \frac{8.1408}{s^4 + 4.4144s^3 + 9.7422s^2 + 12.5952s + 8.1408} \tag{16}$$

$$H_{N=5}(s) = \frac{24.4224}{s^5 + 6.1315s^4 + 18.7979s^3 + 35.6178s^2 + 41.7099s + 24.4224} \tag{17}$$

For the fractional order part, the cut-off frequency gets modified due to the mapping of the radius from *s*-plane to *w*-plane as $\overline{\Omega}_c = \Omega_c^q$, where, $\Omega_c$ is the obtained from conventional third order Butterworth filter design as $\Omega_C = \frac{\Omega_P}{(10^{0.1\alpha_P} - 1)^{\frac{1}{2P}}}$ with $P = 3$. The $\Omega_c$ obtained here corresponds to the *s*-plane. This has to be transformed into the *w*-plane in order to design fractional filters and will thus become $\overline{\Omega}_c = \Omega_c^q$. It has been found that for $P = 3$, $\Omega_c$ is greater than 1. Thus the circle on which the poles lie, will get squeezed on transformation from the *s*-plane to the *w*-plane using Eq. (13).

The nominal fractional part of the BW like filter is found out to be of the form as given in Eq. (18) using the standard notion of designing a third order BW filter but in the transformed *w*-plane.

$$H_{N=0.3}(s) = \frac{1.181}{s^{0.5} + 1.0338s^{0.4} + 1.0688s^{0.3} + 1.105s^{0.2} + 1.1424s^{0.1} + 1.181} \tag{18}$$

The aimed filter design of 4.3 order would be the cascaded version of two filters derived in Eq. (16) and Eq. (18). It is interesting to note that all the poles of the fractional and integer parts in Eq. (16) and Eq. (18) are designed in such a way so as to ensure the same cut off frequency in *s*-plane and *w*-plane by meeting the same set of design criterion.

The upper and lower bounds (i.e. the integer orders) are plotted next in Figure 4 along with the exact FBWF design of order 4.3. The amplitude and phase of the proposed BW like filter asymptotically lie between that corresponding to the design with the upper and lower bounds in the high frequency regions. Figure 4 shows that the proposed FBWF asymptotically falls between the nearest upper and lower integer order Butterworth filter design. Though at lower frequencies the magnitude and phase response of the FBWF of order 4.3 is low compared to the respective integer order cases but for higher frequencies they follow a trajectory which lies between 4$^{th}$ and 5$^{th}$ order classical BW filters. As observed in the magnitude squared response in Figure 5, the FBWF's dc gain seems to be different from the integer order filters. This peculiar behavior is observed due to the presence of ultra-damped and hyper-damped poles in the filter transfer functions where the time response tracking of an input step command takes infinite time, even though the fractional order transfer function has a dc gain of unity as reported in Table 3. This heavily sluggish time response, in spite of mathematically yielding zero steady-state error (from the final value theorem) appears as a relatively low gain at low frequencies.

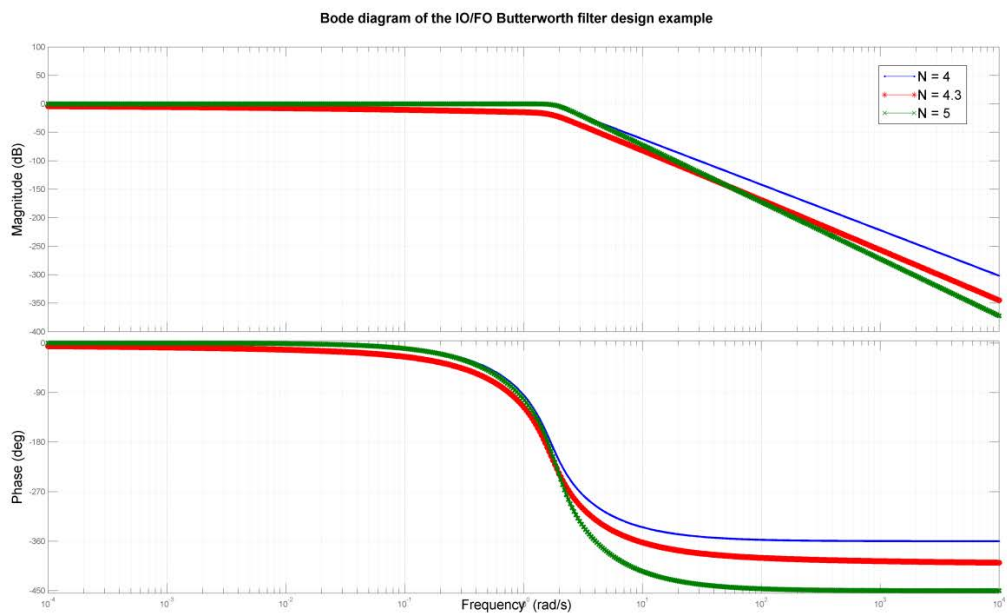

Figure 4: Comparison of the Bode diagrams for the integer order BW filter and the FBW like filter.

### 5.2 Few discussions on the new design technique

Our simulation result is optimum in the sense that we designed a filter of the desired order which exactly satisfies the design specifications and neither under-satisfies nor over-satisfies it. In order to show the effect of rounding off the recommended order for Butterworth filter design by the closest integer values just higher and lower than the exact one is shown in the Bode magnitude and magnitude squared plots for the filter's frequency response. This paper proposes a new type of fractional order design and hence advances theoretical aspect of solving the problem. The choice of the filter orders can be done in future works by minimizing some performance objective and using intelligent optimization algorithms as done in [27].

Also, after obtaining the FBWF, each FO operator can be approximated using well-established analog or digital realization techniques [2]. Since the main goal of the FBWF design is to match the magnitude response, the well-known Charef's method [28] may be easily applied since this technique matches the magnitude roll-off with small piecewise linear operators. For digital realization the FO filter may be expanded using Continued Fraction Expansion (CFE) and Power Series Expansion (PSE) after discretizing the filter using Euler, Tustin or Al-Alaoui rule [29]. In the present paper, the attempt was to have analog domain realization of FBWF. We adopted the approach of designing in $w$-plane and then retrofitting into $s$-plane. The $z$-plane discretization is a further development, where the fractional order theory of stability has to be extended, in a new way. This discretization, in *z*-plane, will not be so direct-vis-à-vis *w*-plane and needs further exploration. Also, as an extension of the present work, the analog FBWF can be transformed to design high-pass, band-pass and band-stop filters from the low-pass design, as done conventionally, since the fractional order operators are linear in nature.

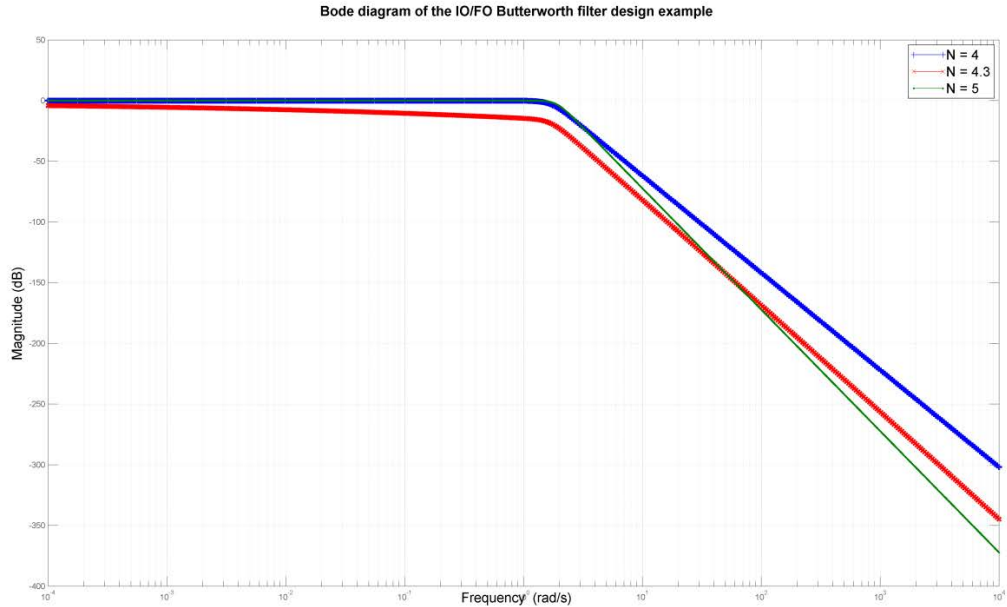

**Figure 5: Magnitude squared function for the IO/BW filter design example.**

## 6. Conclusion

A new kind of fractional Butterworth like filter has been designed using the consideration of poles lying on a circle in the transformed *w*-plane and also using concept of stability in complex *w*-plane. The obtained FBWFs are analyzed from the nature of poles, as well as in the time and frequency domain. The new fractional filter design technique is capable of meeting the design specifications in an exact manner rather than under or over satisfying it. This has been possible by making the filter's order exactly as per the design requirements without rounding-off to the nearest upper or lower integer. A practical design example has been given to show the practicability of the proposed design technique. Future scope of research can be directed towards extending the present concept for discrete domain filtering techniques which has higher flexibility of realization using digital signal processors. Also, the concept can be extended for other class of FO IIR filters and optimization of the filter orders with optimization algorithms as done in [27] can also be done in future.